\documentclass{appolb}
\usepackage{epsfig}
\usepackage{wrapfig}

\newcommand{\del}{$\Delta$}
\newcommand{\epem}{e$^+$e$^-$}
\newcommand{\epep}{e$^+$e$^+$}
\newcommand{\emem}{e$^-$e$^-$}
\newcommand{\piz}{$\pi^0$}
\begin{document}
% \eqsec  % uncomment this line to get equations numbered by (sec.num)
\title{Study of elementary reactions with the HADES dielectron spectrometer 
\thanks{Presented at XXXI Mazurian Lakes conference on Physics}}
% you can use '\\' to break lines

\author{B. Ramstein 
\address{Institut de Physique Nucl\'eaire, CNRS/IN2P3-Universit\'e Paris Sud, F-91406 Orsay Cedex, France}
\and {\scriptsize
G. Agakichiev$^{\,8}$, 
C.~Agodi$^{\,1}$,
A.~Balanda$^{\,3,e}$,
G.~Bellia$^{\,1,a}$,
D.~Belver$^{\,15}$,
A.~Belyaev$^{\,6}$,
A.~Blanco$^{\,2}$,
M.~B\"{o}hmer$^{\,11}$,
J.~L.~Boyard$^{\,13}$,
P.~Braun-Munzinger$^{\,4}$
% ...
}
\address{{\tiny
$^1$Istituto Nazionale di Fisica Nucleare - Laboratori Nazionali del Sud, 95125~Catania, Italy,\\
$^{2}$LIP-Laborat\'{o}rio de Instrumenta\c{c}\~{a}o e F\'{\i}sica Experimental de Part\'{\i}culas , 3004-516~Coimbra, Portugal\\
$^3$Smoluchowski Institute of Physics, Jagiellonian University of Cracow, 30-059~Krak\'{o}w, Poland,\\
$^4$GSI Helmholtzzentrum f\"{u}r Schwerionenforschung, 64291~Darmstadt, Germany,\\
% ...
$^6$Joint Institute of Nuclear Research, 141980~Dubna, Russia,\\
% ...
$^8$II.Physikalisches Institut, Justus Liebig Universit\"{a}t Giessen, 35392~Giessen, Germany,\\
% ...
$^{11}$Physik Department E12, Technische Universit\"{a}t M\"{u}nchen, 85748~M\"{u}nchen, Germany,\\
% ...
$^{13}$Institut de Physique Nucl\'{e}aire , CNRS/IN2P3 - Universit\'{e} Paris Sud, F-91406~Orsay Cedex, France,\\
% ...
$^{15}$Dep. de F\'{\i}sica de Part\'{\i}culas, Univ. de Santiago de Compostela, 15706~Santiago de Compostela, Spain,\\
% ...
$^a$Also at Dipartimento di Fisica e Astronomia, Universit\`{a} di Catania, 95125~Catania, Italy,\\
% ...
$^c$Also at Technische Universit\"{a}t Dresden, 01062~Dresden, Germany,\\
% ...
$^e$Also at Panstwowa Wyzsza Szkola Zawodowa , 33-300~Nowy Sacz, Poland.}
}}
\maketitle
\begin{abstract}
Results obtained with the HADES dielectron spectrometer at GSI are discussed, with emphasis on dilepton production in elementary reactions. 
\end{abstract}
  
\section{Introduction}
The main objective of the High-Acceptance di-Electron Spectrometer at GSI is the study of in-medium modifications of $\rho$ and $\omega$ vector mesons  in hot and/or dense baryonic matter.  Despite the challenging instrumental requirements, the dilepton probe provides the most direct information on the hadronic matter. Being complementary to the ones performed at higher energy facilities (SPS,RHIC) or looking for effects at normal density with photon or proton beams (JLab, KEK), the HADES experiments explore the 1-2 AGeV energy domain, where moderate temperatures (T$ <$ 100 MeV) and baryonic densities up to 3 times the normal nuclear matter density can be achieved, with expected sizeable modifications of $\rho$ and $\omega$ meson spectral functions.  In contrast to reactions at ultrarelativistic energies, the multiplicity of produced pions per participant remains quite small, of the order of 10$\%$. This presents  major advantages, since the  main source of combinatorial background is the conversion of photons from $\pi^0 \rightarrow \gamma\gamma$ or $\pi^0 \rightarrow \gamma e^+e^-$.\par Another specificity of the SIS-18 energy regime is the important role played by baryonic resonances. Due to the very long life-time (15 fm/c) of the dense hadronic matter phase, the resonance can propagate and regenerate and the modification of its spectral function inside the baryonic medium is therefore an important issue for transport model calculations. \par
The $\Delta(1232)$ resonance, which is responsible for a dominant  part of the pion production, is the most copiously produced, but as the incident energy increases, higher lying resonances will play an increasing role. While all of them contribute to pion production, the    N(1535) for example is important  for the $\eta$ production and  the  N(1520), $\Delta(1620)$ and others for the $\rho$ production. Through  the direct  dilepton decay  ($\rho / \omega \rightarrow $ \epem ) or Dalitz decay ($\pi^0 / \eta  \rightarrow \gamma$\epem\ or $\omega \rightarrow \pi^0 $\epem ) modes of these mesons,  the baryonic resonances therefore play a crucial role in dilepton emission. They are also expected to contribute  directly to dilepton emission via their own Dalitz decay modes. For example, the $\Delta(1232)$ should present a Dalitz decay  (\eg $\Delta \rightarrow N e^+e^-$) branching ratio of 4.2 10$^{-5}$, according to QED calculations. 
As it has never been measured up to now, the experimental study of this decay mode is an experimental challenge. In addition, the \del\ Dalitz decay process is in principle sensitive to the electromagnetic structure of the  N-$\Delta$ transition and the kinematics is suited to test the Vector Dominance Models.

On the other hand, another important dilepton source in this energy range is the nucleon-nucleon Bremsstrahlung NN$\rightarrow$NN\epem , which adds coherently to the \del\ Dalitz decay.

Section 2 will show how the first results of HADES in heavy-ion experiments motivate the study of elementary pp and pd reactions. In sect.~3, the \del\ Dalitz decay and NN virtual bremsstrahlung are discussed. Section~4  is devoted  to the description of the experimental set-up and  results from inclusive pp and quasi-free pn reactions. Exclusive measurements in pp collisions and perspectives of pion beam experiments are presented in sect.~5 and 6, respectively. 
\section{First results from heavy-ion experiments}
The  first results from the HADES collaboration for the $^{12}$C+$^{12}$C reaction at 1 and 2 AGeV \cite{Agakishiev06_CC2GeV,Agakishiev07_CC1GeV} have marked an important turning point. In these reactions, the dilepton production shows an excess in the intermediate mass range 0.15-0.6 GeV/c$^2$ with respect to the long-lived source contribution, which is mainly due to the $\eta$ Dalitz decay and is well constrained by experimental measurements.  Such a dilepton excess had been already observed, more than 10 years ago,  by  the DiLepton Spectrometer (DLS) experiment  at Berkeley \cite{Porter97} in the  $^{12}$C+$^{12}$C reaction at 1 AGeV and remained unexplained over years, the situation being known as the "DLS Puzzle". Taking into account   the much smaller DLS acceptance,  a direct comparison of the two data sets was performed, showing very good agreement \cite{Agakishiev07_CC1GeV}. This confirmation of the DLS controversial results triggered new transport model calculations \cite{Bratko08,Thomere07Santini08Schmidt09Barz09} which are now able to reproduce the dilepton spectra measured in the $^{12}$C+$^{12}$C reactions   by DLS and HADES. In particular, the new HSD results are using the recent bremsstrahlung calculation from \cite{Kaptari06}, which is a factor 2-4 higher than other calculations. According to the authors of \cite{Bratko08}, this provided the solution to the "DLS puzzle". However, this  bremstrahlung prediction is contradicted by other approaches \cite{Shyam03}. In addition,  other transport model calculations reproduce quite well the excess with different relative contributions 
of bremstrahlung and \del\ Dalitz decay processes.
More selective experimental constraints on these specific dilepton sources seem therefore necessary to achieve a satisfactory explanation of the intermediate mass dilepton production in the C+C system. This is even more important for heavier systems, like Ar+KCl, recently investigated by HADES \cite{Krizek09} and  where  medium effects are looked for, since a reliable reference for "vacuum" dilepton production above $\eta$ contribution  is  needed. 
This  motivates the study of the p+p and quasi-free n+p reactions  with HADES experiments at 1.25 GeV, i.e. below the $\eta$ production threshold.\par
 Another unavoidable requirement for interpretation of dilepton spectra in terms of medium effects, is a careful description of the vector meson production, which implies the knowledge of inclusive meson production cross-section. The first results \cite{Krizek09} from the Ar+KCl reaction indeed show that this input might need to be readjusted in transport models. To measure it, the dilepton spectra in the p+p reaction at 3.5 GeV have been analyzed.  The results are still too preliminary, therefore, the focus will be in the following on the analysis of p+p and n+p reactions at 1.25 GeV. Results obtained in exclusive analysis at 2.2 GeV will also be presented in sect.~6.
  
\section{\del\ Dalitz decay and NN Bremsstrahlung}  
\subsection{Different theoretical approaches}
The description of these processes has to combine the  electro\-magnetic vertex, including the electromagnetic structure of the involved ha\-drons, para\-me\-trized by form factors and the nucleon-nucleon interaction. The Soft Photon Approximation (SPA) \cite{Gale87} offers a possible way to take both aspects into account, with a factorization of the photon emission probability and of the strong interaction process. It was found to be in reasonable agreement, at least for the pn case, where this process is the most important, with more complete calculations \cite{Shyam03}. As a consequence, SPA is still widely used in transport model calculations, the Dalitz decay dilepton yield being calculated independantly and  added incoherently.\par
Although the  differential decay width of the \del\ Dalitz decay process derives in principle unambiguously from the QED vertex, different expressions can be found in the litterature, as  stressed in \cite{Krivoruchenko01}. We checked ourselves these calculations and could confirm the expressions of \cite{Krivoruchenko01} and \cite{Zetenyi02}.  Consistent spectra are provided by the other descriptions at the \del\ resonance mass pole. However, taking as an example a mass of 1.480 GeV/c$^2$, corresponding to the kinematical limit for a pp reaction at 1.25 GeV, variations of the differential Dalitz decay width  of 30$\%$ for M$_{ee}$ close to zero to 65$\%$ at M$_{ee}$=0.5 GeV/c$^2$ are observed\cite{Wolf90}.\par
\begin{figure}[th]
\centering
	\hspace{.5cm}
		\includegraphics[width= 5 cm]{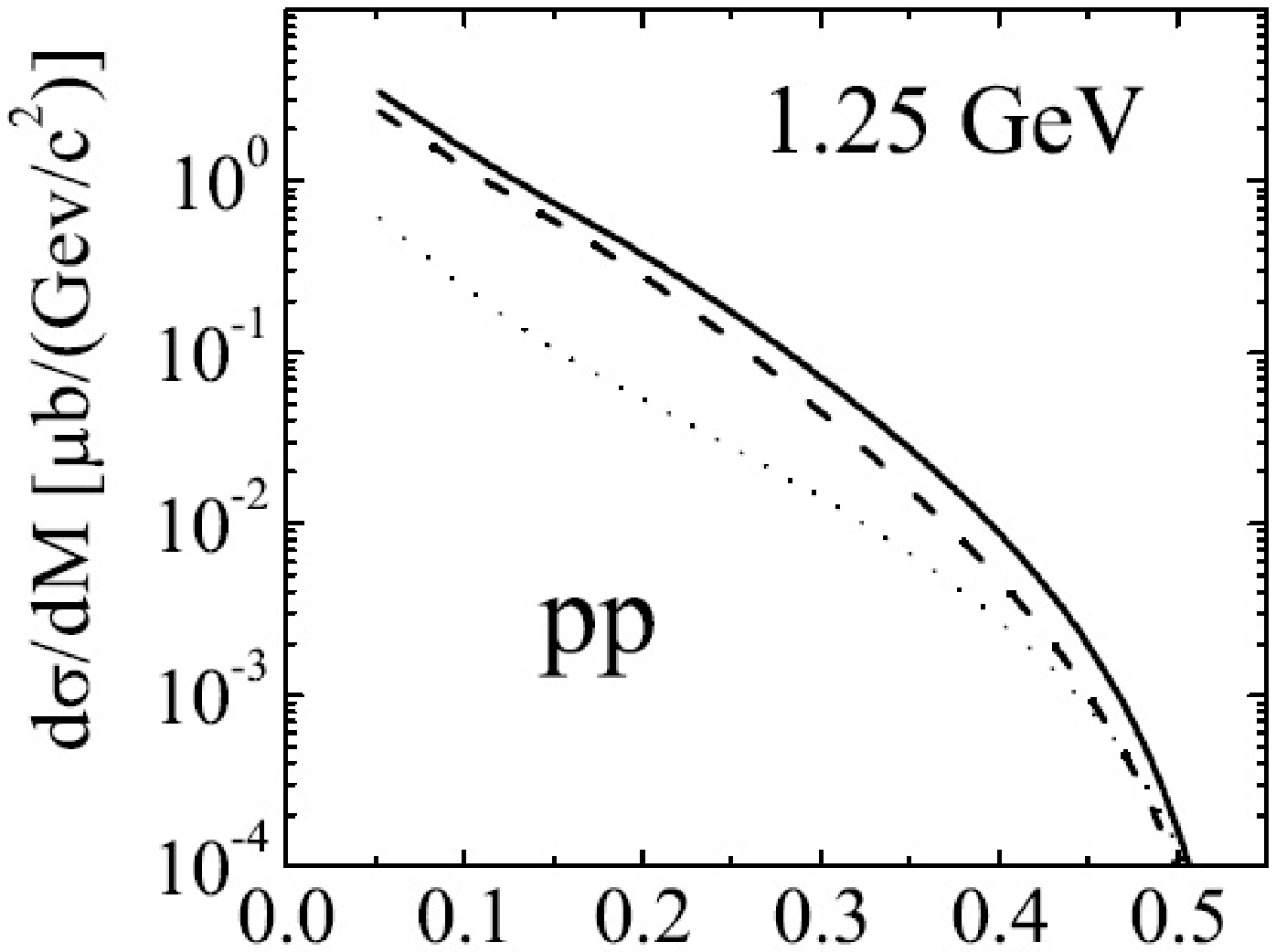}
		\hspace{1cm}
	%	\hfill
		\includegraphics[width= 5 cm]{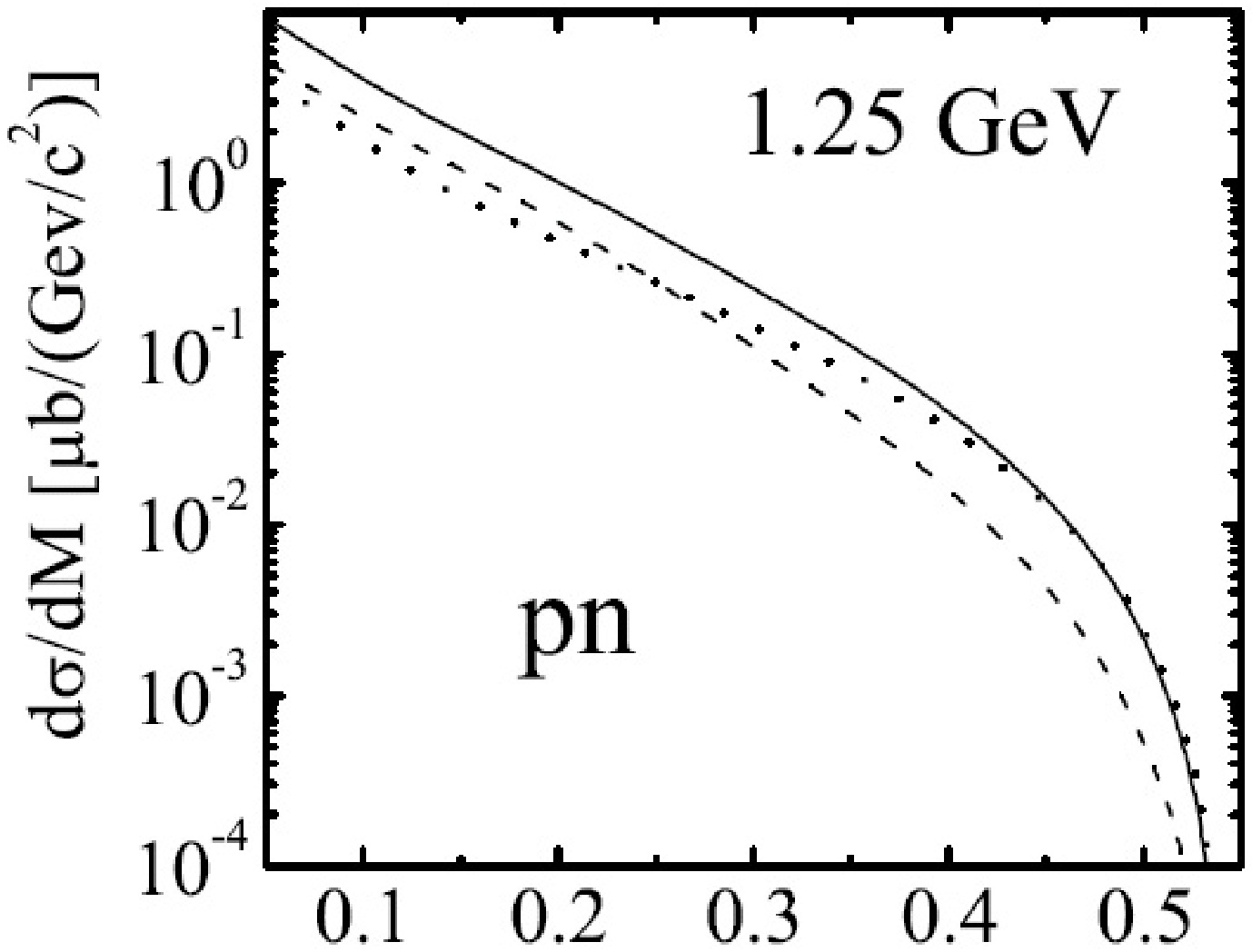}
			\caption{Predictions for the dielectron mass distributions  in the pp (left part) and np (right part)  reactions at 1.25 GeV/nucleon \cite{Kaptari06}. Solid lines: full calculations; dashed lines: \del\ graphs only; dotted lines: nucleon graphs only.}
			\label{fig|kaptari}
			\vspace{-.3 cm}
\end{figure}
A reliable description of these processes can only be accessed in full quantum mechanical and gauge invariant calculations.  Two One Boson Exchange (OBE) calculations \cite{Kaptari06,Shyam09}, which fulfill these requirements were performed recently  and the yields were found  a factor about 2-3 higher in \cite{Kaptari06} than in \cite{Shyam09}  for both pp and pn reactions, the second calculation being much closer to the SPA predictions. 
 An experimental check of these predictions is therefore needed to clarify the situation.\par
 As seen on fig.\ref{fig|kaptari} for an incident energy of 1.25 GeV, the \del\ graphs are widely dominant in the case of the pp reaction, except above 400 MeV/c$^2$, while, for the pn reaction, the nucleon graphs are much more important. These qualitative features are common to both models, and show that, by measuring dilepton spectra in pp and np reactions,  a selective sensitivity to these different  graphs can be obtained.  
\subsection{Electromagnetic form-factors}
Further important elements  are the electromagnetic form factors, which are of two types, namely the elastic nucleon form factor and the N-\del\ transition form factors.  In both cases, the electromagnetic vertex  is  time-like, since the four-momentum transfer squared q$^2$, which is equal to the squared dilepton mass, is a positive quantity.
The influence of elastic nucleon form factors taken in Vector Dominance Models (VDM) has been studied in \cite{Kaptari06}. We will here discuss in more details the case of the N-\del\ transition form factors. 
While, for negative four-momentum transfer squared (space-like region), the three N-\del\ transition form factors (G$_E$, G$_M$ and G$_C$ as electric, magnetic and Coulomb form factors respectively) have been measured in pion-  or photo-production experiments  in a quite wide range of q$^{2}$, the time-like region is unexplored. Here,  the q$^2$ dependence can therefore only 
\begin{figure}[t]
\begin{minipage}[h]{7.5 cm} be given by models, constrained by the  fact that the form factors have to be analytical functions of q$^{2}$, and should reproduce the available space-like data. Due to the small q$^2$ values probed by the dilepton production in our reactions, the major requirement is that the  values of the form factors at q$^2$=0 should be in agreement with the values from pion photoproduction experiments (photon point) and correlatively the radiative decay width ($\Gamma(\Delta \rightarrow \gamma N)=0.66$ MeV $\pm$ 0.06 MeV) should be well reproduced, as in \cite{Krivoruchenko01,Zetenyi02}.  However, the kine\-ma\-tical region probed by the \del\ Dalitz decay 
(q$^2<$  0.3 (GeV/c)$^2$ for an incident energy of 1.25 GeV) is of high\hfill \vspace{-.93cm}
 \end{minipage}\hfill
\begin{minipage}[h]{4.5 cm} 
\vspace{.3cm}
\centering
		\includegraphics[width= 4.5cm]{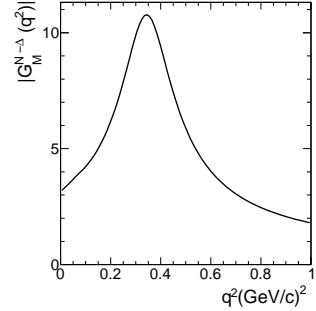}
			\caption{Magnetic N-\del\ transition form-factor squared from the two component quark model \cite{Wan0508}.}
			\label{fig|iachello}
\vspace{-.5cm}
\end{minipage}
%\vspace{-.3cm}
\end{figure}
 interest to check the Vector Dominance.  In such a model, the electromagnetic baryon form factors present structures in the vicinity of the $\rho$  meson mass, which    might be probed by the \del\ Dalitz decay process. 
\subsection{Analysis tool for p+p and quasi-free n+p reactions}
\label{sec|simul}
The new developments of our event generator PLUTO \cite{Froehlich07,Dohrmann09} were exploited in order to  build efficient tools for the interpretation of our data. Two different approaches were followed:\par
The first one is based on the observation that, at an energy of 1.25 GeV/nucleon, pions are mostly produced through intermediate \del\ resonances. In analogy with the description of the \piz\ and \del\ Dalitz decay in transport model calculations, it provides a description of the following channels:
\vspace{-.2 cm}
\begin{eqnarray}
\Delta^+\rightarrow p\pi^0 \rightarrow p\gamma e^+ e^- \ ; \qquad  \Delta^+\rightarrow pe^+e^-\\
\Delta^0\rightarrow n\pi^0 \rightarrow n\gamma e^+ e^- \ ; \qquad \Delta^0\rightarrow n e^+e^-
 \end{eqnarray}
 The cross sections of all the \piz\ and related \del$^+$ and \del$^0$ channels are taken from the resonance model \cite{Teis97}, which describes the existing data \cite{Teis97,Bistricky81}, including as we will see in sect.\ref{exclusive}, the new measurements by HADES in the hadronic channels pp $\rightarrow pp\pi^{0}$. The details of the \del\ production and decay are given in \cite{Froehlich07,Dohrmann09}. For the \del\ Dalitz decay differential width (d$\Gamma$/dM), the expression from \cite{Krivoruchenko01} was adopted, as explained above and two options for the N-\del\ transition form factors are provided: either a constant magnetic form factor (G$_M$=3, in agreement with the photon-point measurements), or the two-component quark-model (fig.\ref{fig|iachello}) \cite{Wan0508}, which is mainly driven by the Vector Dominance in our energy range.    \par
 The second approach aims at a direct comparison with the OBE predictions. Hence, the differential cross sections  (d$\sigma$/dM) provided by the models \cite{Kaptari06,Shyam09} have been parameterized, an isotropic virtual photon emission was further assumed and corrections due to Final State Interaction of the two outgoing nucleons were included. \par
\noindent
 To simulate the quasi-free n+p reaction, the available energy in the center of mass was smeared to include the neutron momentum distribution  in the deuteron using the Paris potential and the energy dependence of the cross-sections was taken into account. When the center of mass energy of the pn system exceeds the $\eta$ threshold, its production is also taken into account, with cross sections taken from existing data \cite{Moskal08}.\par
 The generated events are then filtered by the detector acceptance  in order to compare to the experimental data.  
\subsection{Experimental set-up}
\begin{wrapfigure}{b}{6.cm}
\vspace{-.8 cm}
		\includegraphics[width= 5.8 cm]{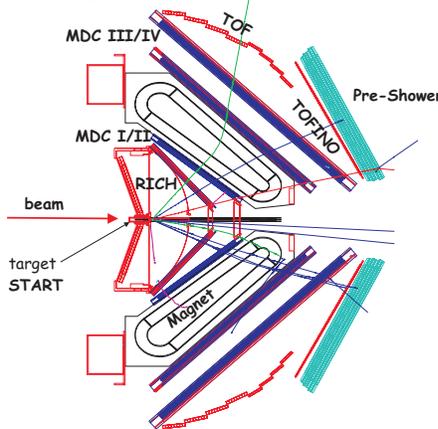}
		\caption{HADES  set-up }
			\label{fig|set-up}
\end{wrapfigure}

The HADES (High Acceptance Dielectron Spectrometer) detector (fig.~\ref{fig|set-up}) consists in 6 identical sectors covering the full azimuthal range and  polar angles  between 18$^{\circ}$ and 85$^{\circ}$, hence providing a lepton pair acceptance of the order of 0.35. A detailed description can be found in \cite{Agakishiev09_techn}, thus only the main features are given here. Momentum measurement derives from the particle trajectory reconstruction using four Mini-Drift Chambers (two 
before and two after the magnetic field zone) providing a position resolution  of about  140 $\mu$m per cell and a measured dilepton invariant mass resolution  of about 2.4$\%$ at the $\omega$ meson mass. A hadron-blind  Ring Imaging CHerenkov detector (RICH), made by a C$_4$F$_{10}$ gas radiator and CsI photocathodes placed around the  target region is used for electron
 identification, together  with  Time Of Flight (TOF/TOFINO) and an electromagnetic pre-shower detector (Pre-Shower).
Particle identification is also provided using the correlation between time-of-flight and momentum for charged pions and protons and   using in addition the width of the time signal in the MDC's for charged kaons.
Time of flight measurements in a Forward Wall scintillator hodoscope (FW) located 7~m downstream the target was used in d+p reactions. It allowed indeed for the detection of forward emitted particles with the characteristics of spectator protons in order to select quasi-free n+p reactions. The first level trigger selects events within a defined charged particle multiplicity range, while the second level trigger corresponds to electron candidates defined by RICH and Pre-Shower/TOF information. 
In the case of the d+p experiment, the first level trigger also requires a coincidence with at least one particle in the FW. 
A 5~cm long liquid-hydrogen target  (1$\%$ interaction probability) and proton and deuteron beams with intensities up to 10$^7$ particles/s were used.        

\subsection{Data analysis}
e$^+$e$^-$ pairs are selected using different criteria to check the track and ring qualities, as well as the identification of the electron and positron.  
The combinatorial background, which  arises from double conversion of  \piz\ decay photons, conversion of the photon emitted in the \piz\ Dalitz decay, or multi-pion decays, was obtained as the arithmetic mean of like-sign \epep\ and \emem\ pairs and was subtracted from the measured \epem\ sample.  The correlated pairs from photon conversion are also removed, using a lower limit  of 9$^{\circ}$ on the opening angle of the pair. Detection and  efficiency corrections, based on GEANT simulations, are also applied, and the final spectra are normalized using the elastic (or quasi-elastic) pp scattering measured simultaneously by HADES. 
The overall normalization error is estimated to be 9$\%$, the systematic error to  about 20$\%$, with a possible smooth invariant mass dependence. In the case of the d+p experiment, a condition on the momentum (1.6 GeV/c $<$ p$_{FW} <$ 2.6 GeV/c) and on the angle ($0.3^{\circ}<\theta_{FW}<6^{\circ}$) of the particle detected in the FW is added.
\subsection{Results and comparison to models }
\begin{figure}[t]
\vspace {0.2 cm}
\centering
		\includegraphics[width= 5cm]{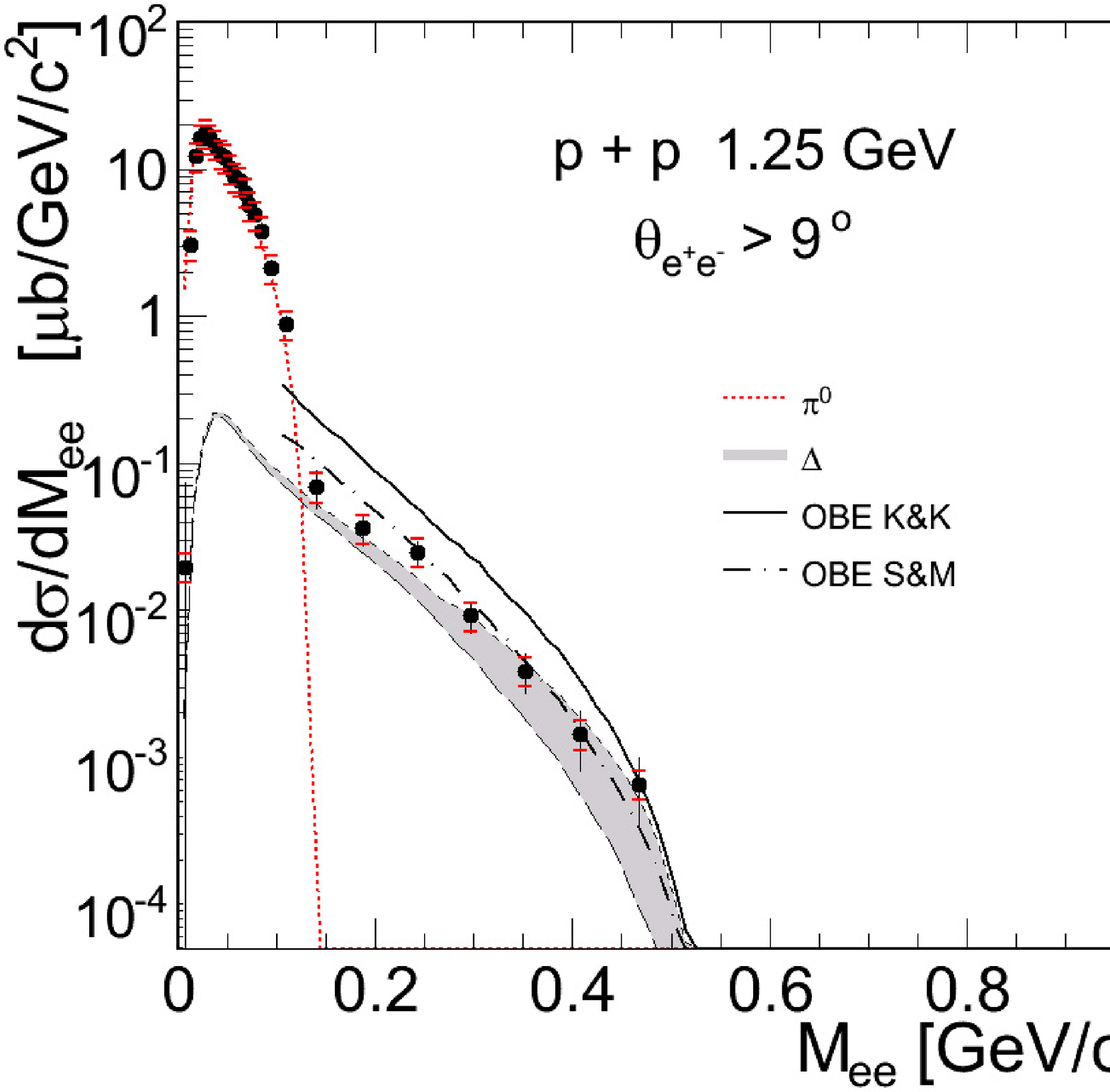}
		\hspace{1cm}
		\includegraphics[width= 5cm]{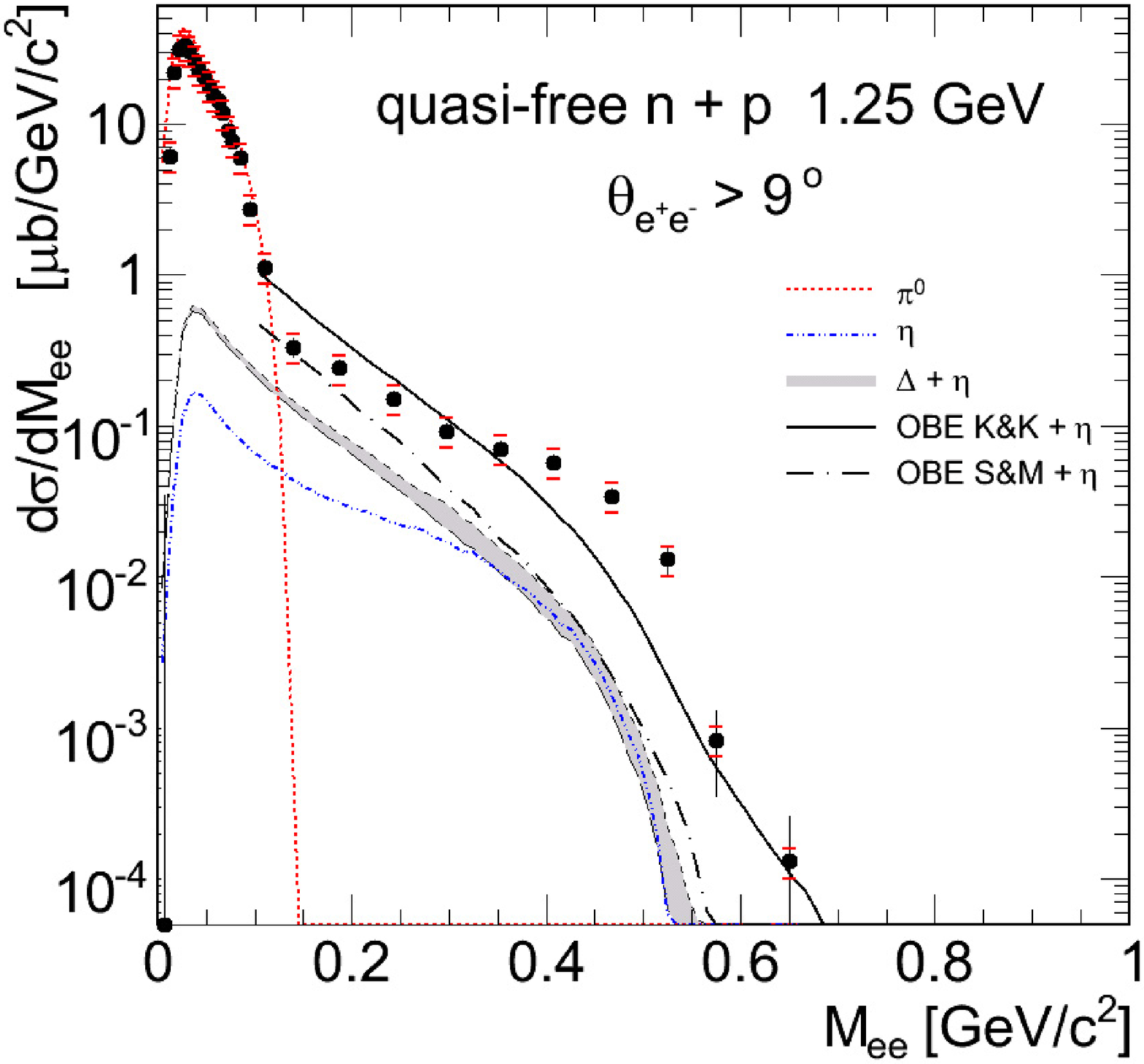}
			\caption{Dielectron mass distribution measured in the pp (left part) and quasi-free np (right part)  reactions at a beam energy of 1.25 GeV/nucleon. The dotted (red on-line) and dashed lines show the contributions of \piz\ and \del\ Dalitz decay, respectively, in simulations using the resonance model. The enhancement due to the N-\del\ transition form factor is shown as the grey area. The dashed and full lines are the results of simulations using the OBE models \cite{Shyam09} and \cite{Kaptari06}, respectively}
			\label{fig|pppn}
			\vspace{-.5 cm}
\end{figure}
Figure \ref{fig|pppn}  shows the dilepton mass spectra measured in the pp and quasi-free np reactions \cite{Agakishiev09_elem} compared to the simulations as described in sect.~\ref{sec|simul}. For both reactions, there is a good agreement between the dilepton yield measured at low invariant masses and the simulation of the \piz\ Dalitz decay, which confirms the normalization and  analysis procedures.  
In the case of the pp reaction, the region of invariant masses larger than 140 MeV/c$^2$ is also well described by the simulation of the \del\ Dalitz decay.  An even better agreement is obtained when the two-component quark model is used instead of the constant magnetic form factor(G$_M$=3), which illustrates  the sensitivity of these data to the electromagnetic structure of the N-\del\ transition.  However, the description of \del\ Dalitz decay in this resonance model is to crude to extract direct information on the time-like N-\del\ transition form factor.   A more accurate description is expected from the OBE models, since they take into account all graphs involving 
intermediate \del\ or nucleons. In these models, constant form factors are used, but defined using different covariants than the usual magnetic, electric and coulomb form factors. This induces a different q$^2$ dependence of the differential width. This effect has again an influence at the high invariant mass end of the spectrum. The predictions of \cite{Shyam09} (shown as dashed line) are indeed in pretty good agreement with the data.  The other OBE model \cite{Kaptari06} (full line) overestimates the data.\par
The shape of the spectrum changes dramatically when going from p+p to n+p interactions. In the  mass region between 0.15 and 0.35 GeV/c$^2$, the yield is about a factor 9 higher in the case of the n+p reaction, while only a factor 2 is expected for the \del\ Dalitz decay contribution due to the isospin factors. The resonance model simulation widely underestimates  the measured dilepton yield. The $\eta$ Dalitz decay contribution is rather small and the inclusion of the N-\del\ transition form factor model does not help either. Nevertheless, this simulation is  missing the nucleon-nucleon bremsstrahlung contribution which is expected to be in the case of the pn system much larger than in the case of pp. The comparison to the OBE exchange models is thus more relevant, but no satisfactory agreement is achieved, even with the model of \cite{Shyam09}, despite its good behaviour in the case of the pp data.   To select more strictly quasi-free reactions, a smaller angular selection ($0.3^{\circ}  < \theta_{FW} < 2 ^{\circ}$) has been applied, with no change in the shape of the invariant mass distribution. No clarification was provided either by
the transverse momentum and rapidity spectra, which present very similar shapes as compared to the p+p reaction.   

These results are also  in agreement with the DLS spectra measured in  pp and pd reactions \cite{Wilson98}, with lower statistics and precision. In the case of the pd reaction, only indirect confirmation  could be obtained through the comparison of the same models, while in the case of the  pp reaction at 1.04 GeV and 1.27 GeV, the direct comparison was possible \cite{Galatyuk09}, showing a very good agreement. The interpretation of the pn dilepton spectra is still the subject of theoretical investigations, related  for example,  to possible   $\rho$ or $\omega$ meson off-shell production by  higher-lying resonances.\par
 These dilepton spectra measured in pp and quasi-free np experiments are used to build a reference spectrum defined by  
 $0.5/\sigma^{NN}_{\pi^0}(d\sigma^{pp}_{ee}/dM_{ee}+ d\sigma^{pn}_{ee}/M_{ee})$, where $\sigma^{NN}_{\pi^0}$ is the mean inclusive \piz\ cross-section in a nucleon-nucleon collision. After subtraction of the $\eta$ contri\-bu\-tions and normalization to the \piz\ multiplicities, the dilepton spectra measured in the C+C systems at 1 and 2 AGeV \cite{Agakishiev09_elem,Galatyuk09,Stroth09} are compatible with this reference spectrum \cite{Stroth09}, which hints to the fact that the excess dilepton yield measured in C+C systems  is due to some additionnal dilepton source already present in the np system. 

\section{Exclusive channels in elementary reactions}
\label{exclusive}
Dedicated exclusive channels can be isolated, by exploiting the capability  of HADES to measure charged hadrons.\par 
For example, the $\pi^0$ and  $\eta$ Dalitz decays could  be studied in pp$\rightarrow $pp\epem $\gamma$ reactions at 2.2~GeV, where  the four  charged particles are detected, the  photon  is reconstructed using  missing  four-momentum and   the meson
\begin{wrapfigure}{b}{7.5 cm}
\vspace{.2cm}
\hspace{0.2 cm}
		\includegraphics[width= 3.3 cm]{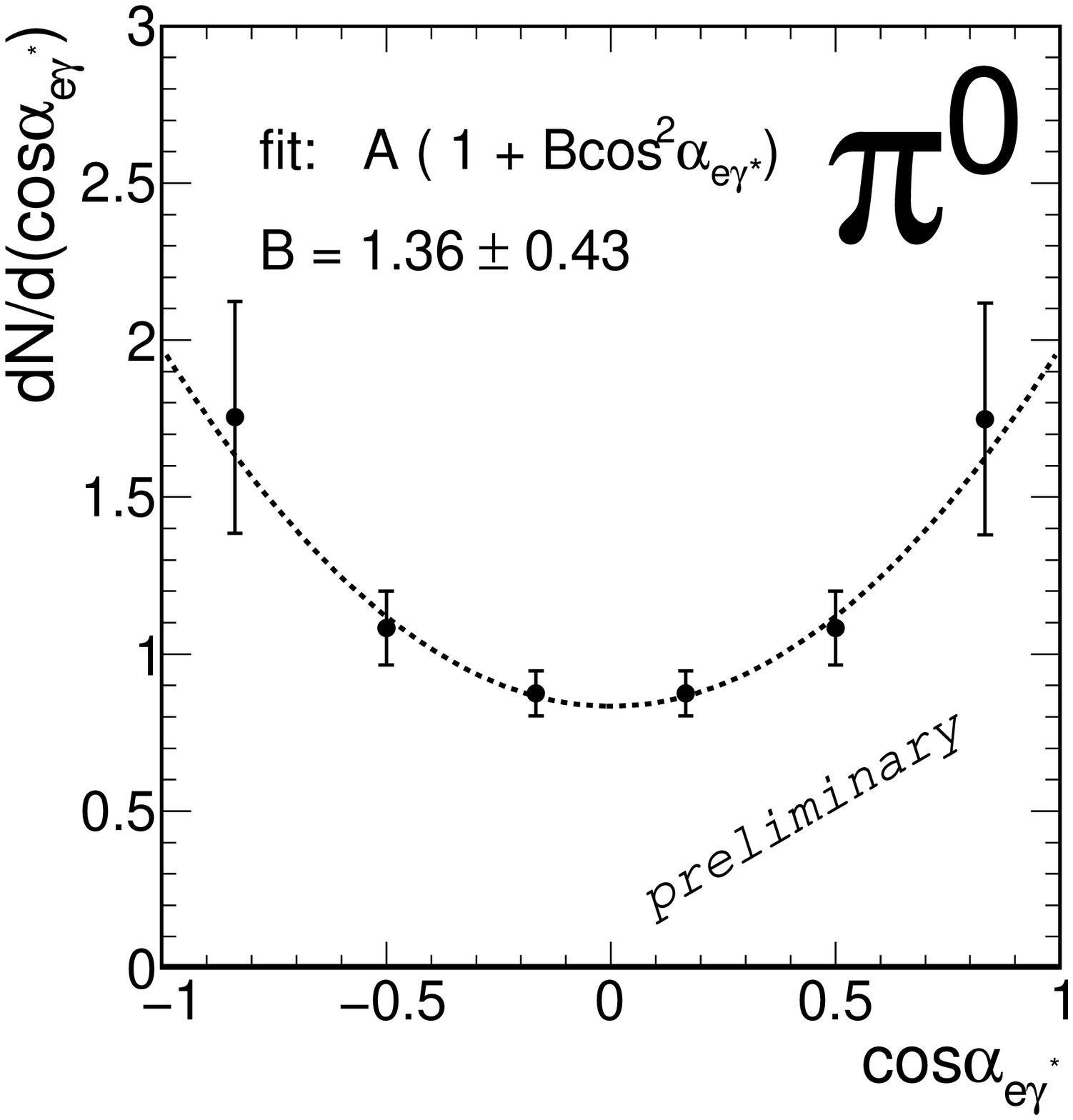}
		\includegraphics[width= 3.3 cm]{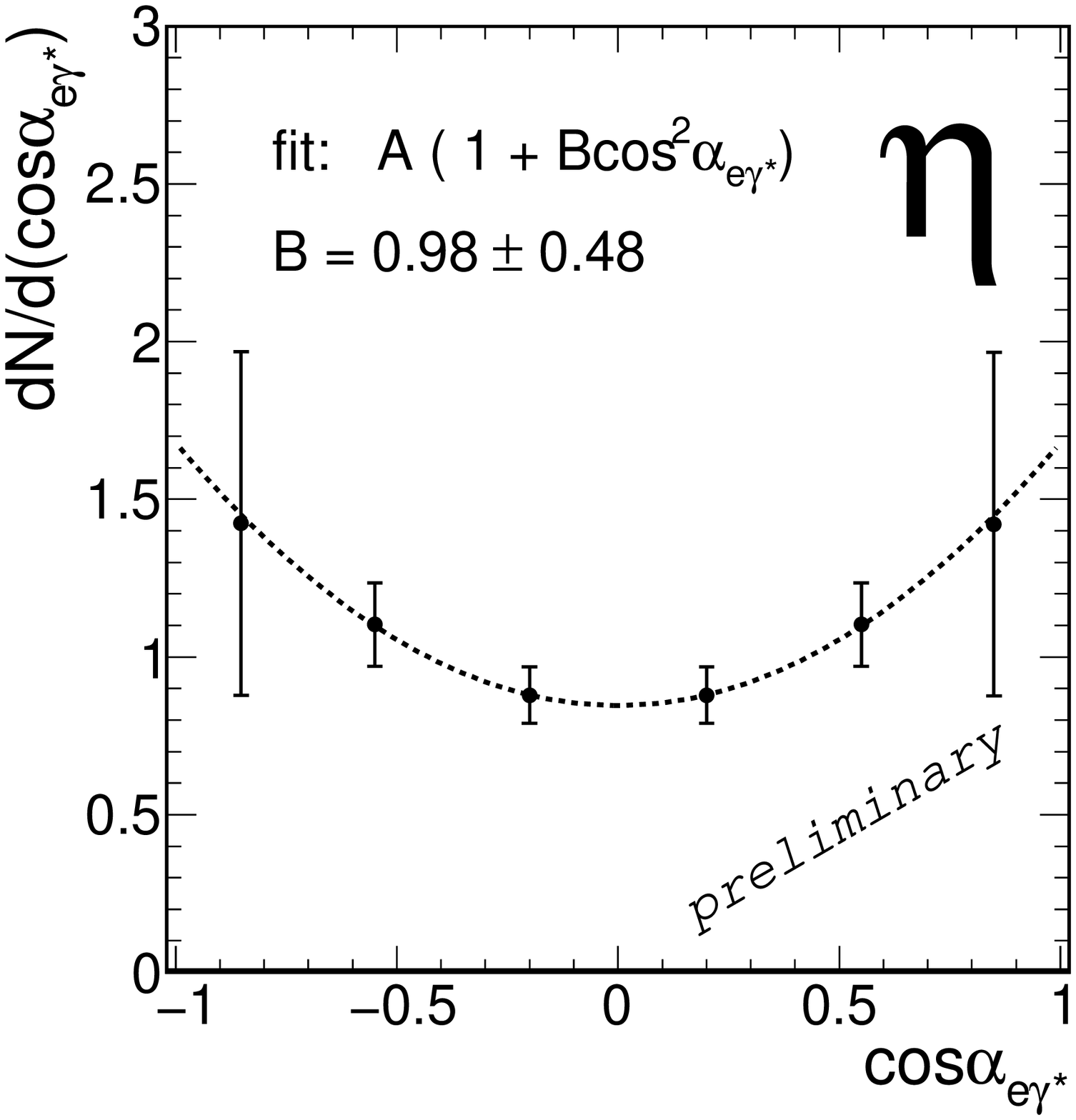}
\vspace{-.2cm}		
	\caption{{\small Helicity angular distributions for \piz\ (left panel) and $\eta$ (right panel) Dalitz decays  reconstructed   from  pp$\rightarrow$pp$\pi^0/\eta\rightarrow$pp\epem $\gamma$ exclusive channels at 2.2~GeV after acceptance and efficiency corrections. }}
	
		\label{fig|helicities}
\end{wrapfigure} 
\noindent   is identified
  by the 2 proton  missing mass.   The helicity angle $\alpha$ is then defined as the angle between the momentum vectors of the lepton in the virtual photon   
 ($\gamma^{\star}$) frame and of the $\gamma^{\star}$ in the decay meson rest frame. After acceptance corrections, these angular distributions are in agreement with a 1+ cos$^2\alpha$ distribution (fig.\ref{fig|helicities}). This is a very nice experimental check of the  trend which is expected from QED, considering that, in the decay of these pseudoscalar mesons, only transverse photons can be produced.
In the pp reaction at 1.25 GeV, the on-going analysis of the pp $\rightarrow$ pp\epem\ channel is expected to bring more detailed information on the \del\ Dalitz decay process and pp Bremsstrahlung, like p\epem\ invariant masses, or lepton helicity angular distribution. \par
Hadronic channels are also intensively studied, since they provide analysis checks, but also new physics results. The detection of both protons from p+p elastic scattering   allows for tracking efficiency and  momentum resolution measurements.
  Moreover,  the exclusive production of unstable particles, which present a known  leptonic or Dalitz decay branching ratio, can be studied both in pp$\rightarrow$ pp\epem X channels, and  in purely hadronic channels, which is suited for a cross-check of the analysis efficiencies, while producing new measurements of the production cross-sections.  This possibility has been exploited in the case of $pp\rightarrow pp\pi^0$ reaction at 1.25 and 2.2 GeV and $pp\rightarrow pp\eta$ reactions at 2.2 GeV.  While the $\eta$ case is still investigated, the exclusive $\pi^0$ production cross-section determined both in hadronic and Dalitz decay channel is found in very good agreement with existing data. \par
\begin{figure}[b]
\vspace{-.5 cm}
		\includegraphics[width= 12cm]{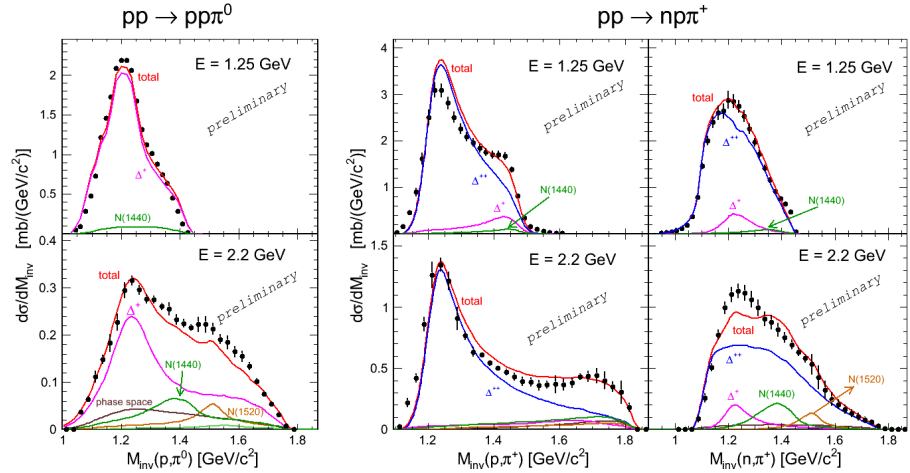}
			\caption{{\small $\pi$ N invariant masses measured in pp$\rightarrow$ pp$\pi ^0$ and pp$\rightarrow$ pn$\pi ^+$ reactions at 1.25 and 2.2 GeV. HADES data (full dots) are compared on an absolute scale to the predictions from the resonance model, with contributions of the following resonances \del$^+$(1232) (pink), \del$^{++}$(1232) (blue), N(1440) (green), N(1520) (light brown) and \del(1600) (light green) and an additionnal small phase space contribution (dark brown).}}
			\label{fig|invariantmass}
\end{figure}
Moreover, the exclusive pp$\rightarrow$pn$\pi^+$ and pp$\rightarrow $pp$\pi^0$ measurements provide very detailed checks of the resonance model used for the analysis of the dilepton spectra \cite{WisniaLiu09}. As shown on fig.\ref{fig|invariantmass}, the yields and invariant mass distributions are in good agreement with the simulations using an event generator based on the resonance model  (see sec.\ref{sec|simul}).  The dominant contribution comes from the  \del\ resonance excitation, but N(1440) and N(1520) play also a significant role at 2.2 GeV. At 1.25 GeV, the relation $\sigma$(pp$\rightarrow$ pn$\pi^+$) = 5 $\sigma$(pp$\rightarrow$ pp$\pi^0$) is fulfilled, as expected from the isospin factors in the different \del\ decay channels.   The angular distributions are also carefully studied, since they carry detailed information on the mechanisms of \del\ resonance excitation beyond the one-pion exchange. \par
\section{Perspectives from pion induced reactions} 
The dilepton spectroscopy in $\pi$ induced experiments on nuclei is proposed in order to study medium effects on $\rho$  and  $\omega$ mesons, with the advantages, with respect to heavy-ion induced reactions, of higher expected effects on   $\omega$ meson and reduced combinatorial background. This would also complement the on-going studies of $\rho / \omega$ production at normal nuclear density in the p+Nb system. Due to the well-known interaction and the possibility of exclusive channel measurements, the reactions on nucleon constitute a unique tool to study  $\omega$  and  $\rho$ production, with a special interest of subthreshold production via the coupling to baryonic resonances. In particular, in \cite{Lutz03Titov01}, a spectacular destructive $\rho$/$\omega$ interference is predicted below the $\omega$ threshold. As these couplings are related to the electromagnetic structure of the resonances, these measurements present a fundamental interest.
 
Strangeness production measurements with pion beam induced reactions is also possible. This includes $\Lambda(1405)$  production in  $\pi$-p reactions at an incident momentum around 1.7 GeV/c,   in medium modifications and K$^-$ absorption in  $\pi^-$-p  and  $\pi^-$-A reactions at 1.7 GeV/c, and K$^0_S$ production in  $\pi^-$+p,  $\pi^-$+C  and   $\pi^-$+Pb at lower energies.\par
From the technical point of view, some developments are still needed to check the feasibility of these experiments. An intensity of 10$^6$ particles/s is needed, which is in principle accessible and fast and thin position sensitive beam detectors are under study to fully reconstruct the  trajectory and momentum of incident pions.

\section{Conclusion}
Recent HADES results have been discussed, with emphasis on the elementary reactions, which allow to build a reference for the heavy-ion measurements, and help to clarify the  controversial problem of contribution of \del\ Dalitz decay and Bremsstrahlung processes.  The interpretation of the inclusive pn spectra  remains however challenging. 
More selective information on the \del\ Dalitz decay and bremsstrahlung processes is expected from the analysis of the exclusive pp$\rightarrow$pp\epem\ reaction.  \par
HADES is currently being upgraded in order to handle more efficiently the higher multiplicities related to the heavier systems, like Ag+Ag and Au+Au. Pion beam experiments also offer interesting perspectives to clarify the role of higher lying resonances.
 Although the specificity of HADES is the dilepton spectroscopy with a large angular acceptance and good precision,  the variety of all these measurements demonstrates the power of HADES as a multipurpose detector.
% \bibliographystyle{../../../../PANDA/prstylong}
%\bibliography{../../../../PANDA/panda}
\end{document}